\documentclass[manuscript,screen]{acmart}
\usepackage{balance}
\usepackage{graphicx}
\usepackage{xspace}
\usepackage{url}
\usepackage{color,colortbl}
\usepackage{tabularx}
\usepackage{framed}
\usepackage{hyperref}
\usepackage{enumitem}
\usepackage{listings}
\usepackage{mathptmx}
\usepackage{dirtytalk}
\usepackage{subcaption}
\usepackage{float}
\usepackage{multirow}
\usepackage{xcolor}
\usepackage{textcomp}

\newboolean{showedits}
\setboolean{showedits}{true} 
\ifthenelse{\boolean{showedits}}
{
	\newcommand{\del}[1]{\textcolor{red}{\sout{#1}}} 
}{
	\newcommand{\del}[1]{} 
	
}

\newboolean{showcomments}
\setboolean{showcomments}{false}
\newcommand{\id}[1]{$-$Id: scgPaper.tex 32478 2010-04-29 09:11:32Z oscar $-$}

\ifthenelse{\boolean{showcomments}}
{\newcommand{\nbc}[3]{
 {\colorbox{#3}{\bfseries\sffamily\scriptsize\textcolor{white}{#1}}}
 {\textcolor{#3}{\sf\small$\blacktriangleright$\textit{#2}$\blacktriangleleft$}}}
 }
{\newcommand{\nbc}[3]{}
 \renewcommand{\del}[1]{} 
 }

\definecolor{ibcolor}{rgb}{0.4,0.6,0.2}
\definecolor{clcolour}{rgb}{0.5,0.7,0.9}
\definecolor{accolour}{rgb}{1,0.5,0}
\definecolor{pmcolour}{rgb}{1,0,0}
\definecolor{amethyst}{rgb}{0.6, 0.4, 0.8}
\newcommand\ms[1]{\nbc{MS}{#1}{ibcolor}}

\newcommand\pu[1]{\nbc{PM}{#1}{pmcolour}}

\usepackage{wasysym}

\definecolor{figurecolor}{RGB}{22,90,220}
\definecolor{citecolor}{RGB}{198,81,19}
\hypersetup{ colorlinks=true, urlcolor=black, linkcolor=figurecolor, citecolor=citecolor, pdfstartview=FitH,
        pdftitle={SoK: Best Practices in Graph Processing Benchmarking},
        pdfauthor={}
        }

\def\Snospace~{\S{}}

\setcopyright{acmlicensed}
\copyrightyear{2018}
\acmYear{2018}
\acmDOI{XXXXXXX.XXXXXXX}

\acmJournal{JACM}
\acmVolume{37}
\acmNumber{4}
\acmArticle{111}
\acmMonth{8}

\begin{document}

\title{SoK: The Faults in our Graph Benchmarks}

\author{Puneet Mehrotra}
\authornote{Equal contribution; order decided by official coin flip}
\affiliation{%
 \institution{Univeristy of British Columbia}
 \country{Canada}}
\email{puneet89@cs.ubc.ca}

\author{Vaastav Anand}
\authornotemark[1]
\authornote{Work done during M.Sc. at University of British Columbia}
\affiliation{}
\affiliation{%
 \institution{Max Planck Institute for Software-Systems (MPI-SWS)}
 \city{Saarbrücken}
 \country{Germany}}
\email{vaastav@mpi-sws.org}

\author{Daniel Margo}
\authornote{Work done as part of Ph.D. at Harvard University.}
\affiliation{%
 \institution{Google}
 \country{USA}}
\email{dmarg@eecs.harvard.edu}

\author{Milad Rezaei Hajidehi}
\affiliation{%
 \institution{Univeristy of British Columbia}
 \country{Canada}}
\email{miladrzh@cs.ubc.ca}

\author{Margo Seltzer}
\affiliation{%
 \institution{Univeristy of British Columbia}
 \country{Canada}}
\email{mseltzer@cs.ubc.ca}

\renewcommand{\shortauthors}{Mehrotra and Anand, et al.}

\begin{abstract}
    Graph-structured data is prevalent in domains such as social networks, financial transactions, brain networks, and protein interactions.
    As a result, the research community has produced new databases and analytics engines to process such data.
    Unfortunately, there is not yet widespread benchmark standardization in graph processing, and the heterogeneity of evaluations found in the literature can lead researchers astray.
    Evaluations frequently ignore datasets' statistical idiosyncrasies, which significantly affect system performance. 
   Scalability studies often use datasets that fit easily in memory on a modest desktop.
    Some studies rely on synthetic graph generators, but these generators produce graphs with unnatural characteristics that also affect performance, producing misleading results.
    Currently, the community has no consistent and principled manner with which to compare systems and provide guidance to developers who wish to select the system most suited to their application.

    We provide three different systematizations of benchmarking practices.
    First, we present a 12-year literary review of graph processing benchmarking, including a summary of the prevalence of specific datasets and benchmarks used in these papers.
    Second, we demonstrate the impact of two statistical properties of datasets that drastically affect benchmark performance. We show how different assignments of IDs to vertices, called \emph{vertex orderings}, dramatically alter benchmark performance due to the caching behavior they induce. We also show the impact of \emph{zero-degree vertices} on the runtime of benchmarks such as breadth-first search and single-source shortest path. 
    We show that these issues can cause performance to change by as much as 38\% on several popular graph processing systems. 
    Finally, we suggest best practices to account for these issues when evaluating graph systems.

\end{abstract}



\received{20 February 2007}
\received[revised]{12 March 2009}
\received[accepted]{5 June 2009}

\maketitle

\section{Introduction}

Graph structured data are used to express complex relationships in social networks\cite{mislove2007measurement}, protein interactions~\cite{proteinNetworks,mason2007graph}, transportation networks~\cite{bast2016route}, and many other domains.
Graphs with billions of nodes and edges are common today~\cite{10.1145/3164135.3164139}, and large-scale industrial applications often require running analytics on trillion-edge datasets~\cite{ching2015one}.
The research community has produced a constant stream of novel graph processing systems over the past decade to address these challenges.

These systems can be designed to distribute their data across a compute cluster\cite{gonzalez2014graphx,gonzalez2012powergraph,low2012distributed} or to run on a single compute node by optimizing the data access patterns - be it from memory~\cite{shun2013ligra,zhu2016gemini} or disk-resident shards~\cite{kyrola2012graphchi,  kim2022blaze,zhu2015gridgraph}.
Together these decisions produce a vast design space with different graph processing systems representing just a single point in this design space.
Application developers thus face the problem of selecting a graph processing system appropriate for their task. 

This challenging task of selecting the appropriate graph system is further exacerbated by the fact that comparing graph processing systems is difficult for several reasons. 

First, there is a lack of robust standardization of benchmarking practices such as which metrics are important, how to measure and report them, and how to do so in a reproducible manner. There have been several attempts at standardization in the High-Performance Computation (HPC) community through the Graph500~\cite{graph500} and GraphChallenge~\cite{Samsi_2018} benchmark definitions. However, these benchmarks are not representative of the diversity of use cases of graph processing and neither do they take into account the immense diversity of datasets that exist in the real world. 
The LDBC Graphalytics~\cite{capotua2015graphalytics} benchmark provides a richer set of datasets and benchmark kernels but remains underutilized in practice. 

Second, most graph processing systems provide their own implementation of popular benchmark kernels. These different implementations mean that when comparing the performance of two systems, it is difficult, if not impossible, to attribute performance differences to the underlying system itself, the algorithm used for a particular kernel, or the quality of the kernel's implementation. For example, Galois~\cite{nguyen2013lightweight} implements four different variations of PageRank~\cite{page1999pagerank}, each with different runtime performance.
Additionally, there are many time-based metrics to compare PageRank: time per iteration, time to run a fixed number of iterations, or time to convergence; different systems use different metrics in their published results.
We found that there is no single reference implementation of triangle counting on \emph{directed} graphs resulting in different graph processing systems reporting different numbers of triangles for the same dataset (\autoref{benchmark_correctness}) . 

Third, there are few large representative graph datasets available for benchmarking, which limits the quality of results being reported.
For example, a distributed graph processing system that claims to be scalable should be evaluated using datasets that are large enough to warrant distribution.
However, we show that this is not true in practice; most systems demonstrate scalability by using popular datasets from the SNAP and KONECT~\cite{snapnets,kunegis2013konect} dataset libraries.
Twitter2010~\cite{kwak2010twitter,twitter2010} is one of the largest, most frequently used datasets, yet in its \emph{uncompressed} state is only 26GB - smaller than the available RAM on a high-end laptop or a modest server.
The dearth of publicly available large datasets poses a significant problem that is frequently addressed via synthetic graph generators.
The RMAT~\cite{chakrabarti2004r} and Kronecker graph generators~\cite{leskovec2005realistic,leskovec2010kronecker} remain the most popular.
However, these generators do not produce graphs that are representative of real world graph datasets, which distorts performance~\cite{anand2020smooth}.

Fourth, statistical properties of datasets produce hidden, uncontrolled effects at various steps of the graph processing pipeline that drastically affects the performance on the systems. We demonstrate the effect of two such properties inherent in the graph structure of datasets - vertex ID assignments (\autoref{sec:isomorphism}) and the presence of zero-degree vertices (\autoref{sec:zero}), each of which can introduce changes in key performance metrics such as runtime and cache
behavior. Changing the vertex ID assignment of a dataset can cause a performance difference of as much as 38\% for the PageRank benchmark on several popular graph processing systems; the presence of isolated vertices can cause a 10x performance boost for benchmarks such as BFS.
\emph{Failure to identify these issues produces results that reflect idiosyncrasies of the experimental setup rather than the fundamental behavior of the systems under evaluation.}

Together, these challenges render system comparisons obscure, at best, and meaningless at worst.
In addition to these challenges, different developers might have different goals for their application: minimizing runtime, running within a memory or cost budget, rapidly ingesting a stream of updates to the graph, etc. \emph{How can we, as researchers, help developers make wise choices?}

To help developers make wise choices about using, building, and benchmarking graph processing systems, we present a systemization of knowledge surrounding the benchmarking of graph processing systems. Specifically, we present the three following contributions.
First, we provide a study of 12 years of published literature on graph processing systems and algorithms that establishes the current status quo for benchmarking these systems.
Second, we present empirical data that illustrates the significance the properties of datasets can have on runtime, using four graph processing systems, the most popular benchmarks, and the most widely used datasets identified in the literature survey. 
Finally, we provide a set of principles (\autoref{sec:best}) to guide evaluation of graph processing systems.
\section{12 Years of Graph System Benchmarking}
\label{sec:metastudy}
\begin{figure*}[hbt]
    \centering
    \includegraphics[height=0.50\textwidth,width=\textwidth,keepaspectratio]{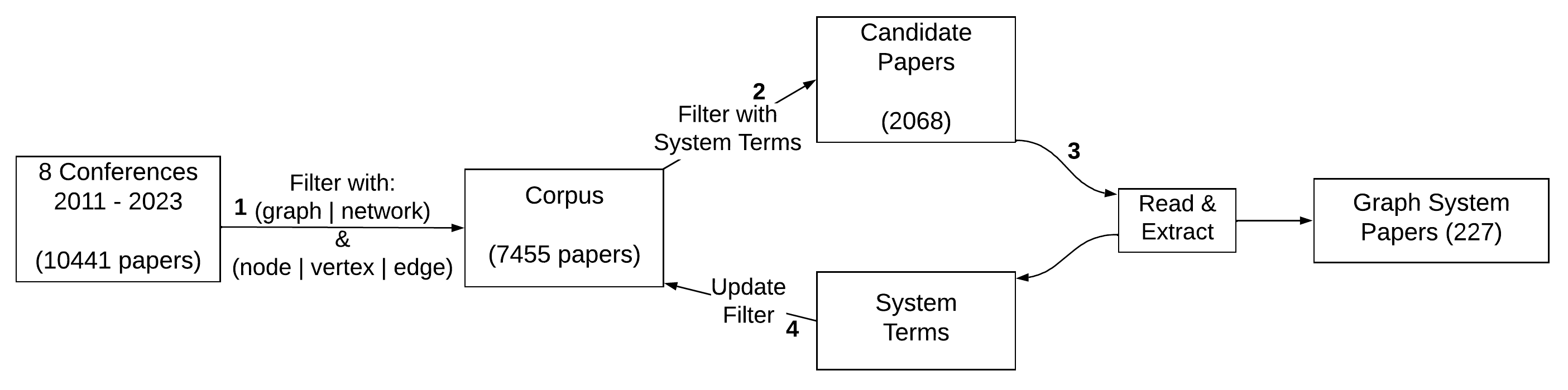}
    \caption{Our procedure to identify contemporary graph system papers. 1.) Our
    corpus consists of papers published at 10 conferences held between 2011 to 2023 and are pruned using common-sense
    graph terms. 2.) We filter our corpus with major graph system names to
    obtain candidate papers. 3.) We read the papers and extract more graph
    system terms from them. 4.) We update the filter with the new terms and
    repeat the process.}
    \label{fig:filter_proc}
\end{figure*}

\begin{table*}
    \centering
    \begin{tabular}{|l|r|r|r|r|r|}
        \hline
        Venue  & Years         & Total & Graph Related & Refers to System &
        Presents System \\ \hline
        GRADES & 2013-22       & 113    & 110           &71              & 22 \\
        \hline
        KDD    & 2011-22       & 3316  & 2236         & 477             & 10 \\
        \hline
        OSDI   & 2012-23       & 368   & 298           & 106              & 9 \\
        \hline
        PODS   & 2011-22       & 359   & 258           & 57              & 0 \\
        \hline
        SIGMOD & 2011-22       & 1976  & 1350           & 456             & 63 \\ 
        \hline
        SOSP   & 2011-21       & 217   & 177          & 56              & 8 \\
        \hline
        VLDB   & 2011-22       & 2574  & 1729          & 505             & 86 \\
        \hline
        NSDI   & 2011-23       & 638   & 541          & 233               & 8 \\
        \hline
        EuroSys & 2011-23      & 485   & 467          & 99             & 16 \\
        \hline
        FAST & 2011-23      & 395   & 289          & 8             & 5 \\
        \hline
        \textbf{Total} & \textbf{2011-23} & \textbf{10441} & \textbf{7455} & \textbf{2068}    & \textbf{227} \\ \hline
        \end{tabular}
    \caption{Venues used in the study and the results of the paper selection
    pipeline.}
    \label{tab:papers}
\end{table*}

Each time we set out to conduct a performance evaluation including
widely used graph processing systems, we encountered contradictory
and confusing results.
Ultimately, we decided to catalog a decade of research
in such systems, which uncovered the host of problems discussed
in the previous section.

\subsection{Methodology}
The volume of graph processing research and its generality raises questions as
to what criteria define a graph systems paper. Our choice of venues in which we
search for papers is similarly ambiguous. 

We resolve this ambiguity by clearly defining our ideal corpus and then
methodically constructing such a corpus as best we could. Ideally, \emph{we want
to include every paper that describes a graph processing system and was published
after 2010's Pregel, which is generally regarded as the first ``vertex
programming'' system and a major motivator for recent research}. A ``graph
processing'' system is an implementation that explicitly operates on
graph-structured data and produces analysis or query results. Therefore, any
relevant paper should explicitly discuss graphs. We impose no generality
requirements on such a system; therefore, an implementation of a
single-purpose graph algorithm is a graph processing system; any graph mining paper describes a graph processing system.
Conversely, a
general-purpose processing framework that ``can'' implement graph algorithms
is not within our scope unless some paper explicitly discusses those graph
algorithms. For example, we include a paper in our corpus that presents new relational
algebra operators for processing graph-structured
data~\cite{all_in_one_presents_system}, while we discard a paper that presents an optimization on Gradient
Descent~\cite{gradient_descent_not_system}. 

We did our best to adopt a consistent method in our paper search, but ultimately
our judgment and expertise do play important roles in the search process.
Inevitably, some distance will exist between our experience and the aggregated
experiences of our readers. This could be construed as a sampling error
or noise; the nature of such ``error'' changes with the expectations of the
reader. Rather, we encourage the reader to read this paper as an autopsy and
diagnosis of some limited, but relatively unbiased, subset of the graph
processing community. 

\subsubsection{Selecting Papers}
We began by downloading the full proceedings from the venues indicated in \autoref{tab:papers} (10441 papers).  We followed this step with a recursive
filtering process, which uses an expanding set of graph systems (our search
phrases), that continues until it produces no new systems. 

In the first step, we filtered the corpus to identify papers that might be
related to graph systems. To do that, we devised the following simple rule as a
first-level filter: a paper might be graph-related if it contains the terms
\emph{graph} or \emph{network} \emph{and} any one of \emph{node}, \emph{vertex}, or
\emph{edge}. This first-level filtering produces an initial set of papers
that are likely related to graph processing. At the end of this stage, we were
left with 7455 papers.

In the second level filtering, we use the reduced corpus from the first
stage, and search for \emph{Giraph, GraphLab, GraphX, PowerGraph, and Pregel} -
which we consider to be the first few important graph systems. We then manually
inspect all the papers containing these seed phrases to find new graph systems.
We add each newly found graph system to the set of seed phrases and iterate
until we stop finding new systems.

This methodology suffers from several flaws - not all systems are named, many
papers discuss the same system or name systems that are outside our initial
corpus, and the selection of papers in the filtering process suffers from the
writers' bias. Therefore, there is no one-to-one mapping between papers and
search terms. We present this metastudy as a review of a limited subset of the
graph processing community. However,
this process, illustrated in \autoref{fig:filter_proc}, identified 2068 papers,
from which we manually identified 227 papers that describe systems or algorithms
that process graph data.

\subsection{Findings}
For each of the 227
papers, we manually determined all the benchmarks and datasets used in its
evaluation. This process is necessarily manual, because there is a significant
entity linkage concern regarding benchmarks and datasets. For example, a paper
might reference a ``KONECT Wikipedia dataset'', but in fact there are numerous
Wikipedia datasets on KONECT and the specific dataset can be identified only by
its vertex and edge counts. \pu{we mention this again in the next sub-section. Drop?}
This is further complicated by variations in the
reported vertex and edge counts of identical datasets - we encounter papers that
reference names that refer to non-specific datasets as well as specific datasets
with varying features. 


\begin{figure}
    \centering
    \includegraphics[width=0.48\textwidth]{"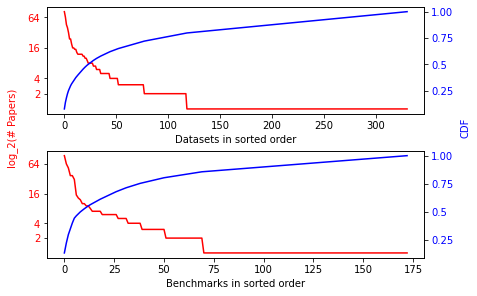"}
    \caption{Frequency of usage of datasets and benchmarks that appear in our 227 paper corpus. The red axis shows the number of papers that use that particular dataset or benchmark. The blue axis shows the CDF of the percentage of the usage of a particular dataset or benchmark out of the total usage of the total usage of all datasets or benchmarks across our paper corpus.}
    \label{fig:dist}
\end{figure}

\begin{figure}
    \centering
    \includegraphics[width=0.48\textwidth]{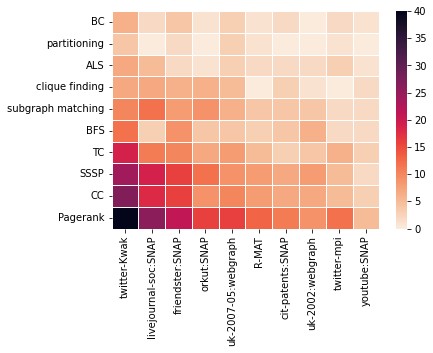}
    \caption{The frequency of co-usage of the top-10 benchmarks and top-10 datasets}
    \label{fig:dist:datasets_benchmark}
\end{figure}

\subsubsection{Datasets}

We identified 331 different datasets used in our corpus.
 However, fewer than 5\% of these datasets have been used in more than 10
papers.
The top 10 datasets account for over 37\% of all datasets
used. \autoref{fig:dist} shows the usage of the various datasets
across research papers. On average, a dataset is used in only three different 
papers.

The most widely used dataset, appearing in more than one-third of the research papers in our corpus,
is Twitter2010~\cite{twitter2010}.
This is problematic in a couple of different ways.
First, papers report different numbers of edges for Twitter2010~\cite{shao2014parallel,hong2013early}.
Second, there are (at least) two different Twitter datasets and some papers in our corpus simply refer to ``the Twitter dataset.'' 
There is a second Twitter dataset from MPI~\cite{icwsm10cha}, which has 10 million more vertices and 200 million more edges than Twitter2010~\cite{twitter2010}.
While this inconsistency might not cause problems if all comparisons are done with the same dataset, it can be problematic to compare systems that do not make their artifacts publicly available.

As can be seen in \autoref{fig:dist:datasets_benchmark}, social network datasets and synthetic graphs (that produce a large social network-like graph) are the dominant graph type used for comparative benchmarking.
The large skew in degree distribution of such graphs is a useful tool to test the partition quality of a distributed graph processing system and examine its load-balancing scheme.
However, not all graphs possess similar degree skew, and not all graph processing systems are designed to be distributed.
There are other statistical properties of a dataset that can determine its utility in stressing different graph processing systems - large diameter, sparsity (number of edges per vertex), average triangle count per vertex, and maximum in- and out-degrees to name a few.
These are often ignored in favor of selecting the largest graph researchers can find.
For instance, Besta et al.~\cite{10.14778/3476249.3476252} identified that Flickr~\cite{FlickrImage} and Livemocha~\cite{LiveMocha} have similar node and edge counts and diameters but vary greatly in the performance characteristics for 4-clique mining due to the large difference in the number of triangles (indicative of clustering property).
A social network such as Livemocha should have a few 4-cliques of friendships, but in a network such as Flickr, where related photos share some metadata (such as location), 4-cliques should be common.
Graph Mining Suite \cite{10.14778/3476249.3476252} presents a better approach to selecting appropriate datasets for benchmarking: emphasizing a rich diversity of dataset origins and selecting them in a way that enhances the represented statistical properties.\pu{we should have a reference to a figure in this paper where they list datasets and *why* they use them in their suite. I think it's valuable.}

Most disturbingly, we found that the Netflix dataset is one of the top 10 most widely used datasets, even though \emph{it is illegal to distribute this dataset and arguably unethical to experiment with it}. 
The dataset contains recommendations by users who were later deanonymized by researchers~\cite{narayanan2008robust}. 
Both an FTC inquiry and a class action lawsuit that Netflix settled in 2010~\cite{singel_2017} caused Netflix to cancel further competitions and withdraw the dataset, whose license clearly states that  it cannot be redistributed without permission. 
Nevertheless, between 2010 and 2023, nine papers in our
corpus had used this dataset. \emph{This practice must stop.}

Less legally and ethically troubling, but more technically troubling, is the
fact that we found inconsistencies in the number of vertices and edges reported
for the datasets across the papers. \emph{This makes interpreting results tricky as
one can't be sure if two papers use the same dataset.}

This issue of inconsistently sized datasets is exacerbated by the practice
of simply naming datasets without reference.
In our corpus we found twelve
different variants of ``the Wikipedia dataset''; we were unable to locate five
of them even after extensive web searching. Four of the remaining Wikipedia
datasets simply cited KONECT as the source of the dataset but KONECT has 37
different ``Wikipedia datasets''. Similar issues arose with datasets named
uk-20XX; some papers cite them only as ``the uk dataset''. In our corpus, we
found at least four different uk datasets being used. On further investigation, we
found that Laboratory for Web Algorithms~\cite{BoVWFI,BRSLLP,labwebalgorithmics}
is the source of the uk datasets and has 14 different uk datasets.

\subsubsection{Synthetic Graph Generators}

\begin{figure}
    \centering
    \includegraphics[width=0.48\textwidth]{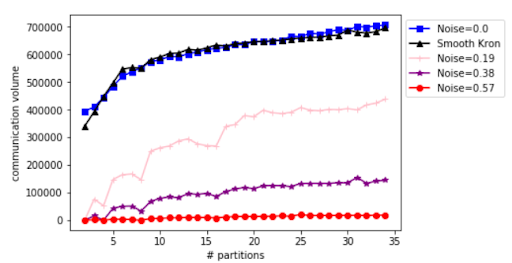}
    \caption{Partitionability of graphs generated by Smooth Kronecker compared to graphs generated by Noisy Kronecker generators}
    \label{fig:dist:smoothkron}
\end{figure}

Real graph datasets are usually small in size and often fit in memory. In the absence of readily-available truly large graphs, researchers rely on synthetic graph generators to such an extent that
graph datasets generated from RMAT or Kronecker graph generators~\cite{shun2013ligra,chakrabarti2004r,leskovec2005realistic,leskovec2010kronecker,park2017trilliong,wsvr} are the
third most commonly used source for datasets.

Despite their popularity, default Kronecker generators are not well-suited for
benchmarking graph processing systems as these generators produce
graphs with unrealistic node degree distributions inducing a ``combing effect''~\cite{anand2020smooth}.
To rectify this, researchers developed Noisy Kronecker models that add
uniform random noise to the graph generation process to fix the degree distribution problem~\cite{seshadhri2013depth}.
But graphs generated by Noisy Kronecker models generate graphs
that are more easily partitionable than the original graph the generator models.
Smooth Kronecker~\cite{anand2020smooth} is a synthetic graph generator that
fixes the ``combing effect'' without adding any noise in the generation process.
Additionally, \autoref{fig:dist:smoothkron} shows that there is no difference in the partitionability
of the graphs generated by the Smooth Kronecker model
and the original graphs.
The Smooth Kronecker graph generator produces graph
datasets that are better suited for benchmarking graph processing systems.

However, none of these graph generators were designed to generate graph datasets that evolve over time.
With modern cloud applications increasingly operating on real-time evolving data, dynamic graph algorithms
have been developed that operate on dynamic graphs. To the best of our knowledge, there exists only one graph generator,
DyGraph~\cite{mccrabb2022dygraph}\footnote{Available at \url{https://adacenter.org/dygraph}} that produces graph datasets for benchmarking dynamic graph algorithms.

\subsubsection{Benchmarks}

We identified 173 unique benchmarks in our corpus;
only $\sim$5\% (4.6\%) of the benchmarks have been used in
more than 10 papers and the 10 most frequently used benchmarks account
for 49\% of all benchmark usage.
\autoref{fig:dist} shows the
distribution of the various benchmarks used by the papers in our corpus.
On average, a benchmark is used in only three different papers;
only a select few are widely used.
This use of bespoke benchmarks
makes it difficult to understand precisely what is being measured or gain any
intuition about the system under study.

PageRank, used in over 54\% of the papers, was the most popular benchmark.
However, there are two fundamentally different ways to compute page rank.
Some systems
compute page rank at each vertex for a fixed number of iterations, while others
compute page rank until the PageRank values converge.
In our corpus, over 55\% of the PageRank usages report the elapsed time
for either a single iteration or for a small, fixed number (usually 10 or 20)
of iterations.
About 20\% of the papers
using PageRank as a benchmark reported the time it took for PageRank to
converge,  while the remaining 25\% just report ``runtime'' without any
further specification about whether the benchmark runs for some fixed
iteration count or until convergence. 

Even comparing results that all measure time for a specified number of
iterations is challenging.
Usually, each iteration updates each vertex exactly once; however, some newer
systems use asynchronous methods that run on every vertex as often as new data
becomes available, or they prioritize vertices that are far from convergence. In
contrast to the fixed iteration setup, an asynchronous implementation typically
runs until the page ranks for vertices are within a target tolerance of
convergence. Synchronous PageRank implementations are so common in contemporary
graph systems evaluation that it discourages comparisons to asynchronous
implementations of PageRank.
For example, PowerGraph (2012)~\cite{gonzalez2012powergraph}
was one of the first vertex programming systems to support asynchronous
execution. However, in their comparative evaluation they only benchmark synchronous PageRank,
because prior works measured and published ``per-iteration'' runtimes that are
defined only for synchronous implementations.

The authors of GraphMat (2015)~\cite{graphmat}, an iterative Sparse Matrix
Vector Multiplication (SPMV) system, use Synchronous PageRank to
compare to PowerGraph and Galois, concluding that GraphMat is 2.6X faster per-iteration
than Galois. This factor is comparable to the difference between the best
synchronous and asynchronous implementations in Galois. Galois outperforms
GraphMat on every end-to-end runtime measure but underperforms on every
per-iteration measure, from which GraphMat's authors conclude that, on average,
they outperform Galois; this claim is almost entirely due to synchronous
PageRank results. To be clear, GraphMat's evaluation followed
conventional best practices and was consistent with prior work. Our concern is
that \emph{consistency with prior evaluations has brought the field to a state
where the most popular comparative benchmark discriminates against alternative
and arguably superior solutions.}

Such inconsistencies are also visible in recent papers on dynamic
graph processing systems.
These systems support fast analytic performance in the presence of
modifications to the graph \cite{besta2021practice}.
For example, LLAMA~\cite{macko2015llama}, GraphOne~\cite{kumar2020graphone}, and
Teseo~\cite{de2021teseo} are three such systems, published in 2015, 2020, and
2021 respectively.
The GraphOne paper claimed to be 4x faster
than LLAMA in PageRank workloads, however, the Teseo paper (published
after the other two) showed that LLAMA was consistently
faster than GraphOne in PageRank (by up to 3x).
Such inconsistencies
demonstrate that various experimental features such as
the degree of parallelism, hyperparameter settings,
dataset characteristics (e.g., the size and topology of the graph),
or implementation of algorithms can significantly affect performance.

Without being fully aware of what workload characteristics are important
and how systems perform with respect to them, it is impossible for
practitioners to identify the right system.
Hence, there is a need for standard benchmarking specifications that,
at a minimum, explicitly state all the relevant features, and
better yet, evaluate systems under a range of them.



\subsubsection{Co-occurence of datasets and benchmarks}
\autoref{fig:dist:datasets_benchmark} shows the co-occurrence frequency of
the most commonly used top eleven datasets and benchmarks. Using
PageRank on the Twitter2010~\cite{twitter2010} dataset is, by far, the most popular choice in our
corpus. This substantiates the claim made by the Naiad
authors in 2013 that ``several systems for iterative graph computation have
adopted the computation of PageRank on a Twitter follower graph as a standard
benchmark''~\cite{murray2013naiad}. Unfortunately, the publishers of the
Twitter2010 dataset have reported that ``Twitter2010's follower-following topology is
substantially different from other social networks in terms of its non
power-law degree distribution, short-effective diameter, and lack of
reciprocity''~\cite{kwak2010twitter}. This makes the results obtained for running
various benchmarks on PageRank and Connected Components not generalizable to
other social network datasets.

The popularity of the Single-Source Shortest Path (SSSP) benchmark with
unweighted datasets such as Twitter2010~\cite{twitter2010} and soc-LiveJournal~\cite{socLiveJournal} is equally surprising as
SSSP requires edge-weights. When using this benchmark with these data sets,
researchers produce weighted datasets from unweighted datasets \emph{by
assigning random weights to the edges}. Details of the randomization and the
weights are never documented. This make it impossible to reproduce the results for SSSP benchmarks. 

Triangle Counting is another popular benchmark used in 16\% of our papers.
Triangle Counting is well-defined on undirected graphs but is ambiguous for
directed graphs. In a directed graph, a triangle counting kernel can count either cycle triangles or trust triangles. 
Given that the top 10 most popular real graph datasets are directed, using a triangle counting kernel intended for undirected
graphs on directed graph datasets is an example of a mismatched dataset-benchmark combination.

\subsubsection{Conference Conformance Bias}

\begin{figure}[htbp]
\centering
\begin{subfigure}[b]{0.45\linewidth}
  \centering
  \includegraphics[width=\linewidth]{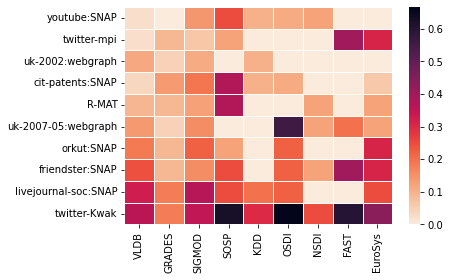}
  \caption{}
  \label{fig:dist:dataset_conf}
\end{subfigure}
\begin{subfigure}[b]{0.45\linewidth}
  \centering
  \includegraphics[width=\linewidth]{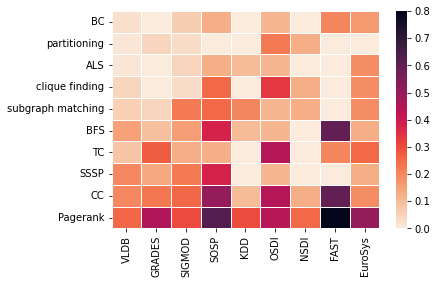}
  \caption{}
  \label{fig:dist:benchmark_conf}
\end{subfigure}
\caption{(a) shows the use of the top-10 datasets across conferences; (b) shows the use of the top-10
    benchmarks across conferences. We omit PODS as there is no paper from PODS in our corpus.}
\label{fig:side_by_side}
\end{figure}


\autoref{fig:dist:dataset_conf} and \autoref{fig:dist:benchmark_conf} show
dataset and benchmark use across conferences. Each box represents the
percentage of graph papers at that conference that used a particular
dataset/benchmark. A co-occurrence correlation exists between the papers
accepted to a conference and the benchmarks used in that work. As can be seen in
\autoref{fig:dist:benchmark_conf}, for example, more than 50\% of the graph papers at
SOSP, FAST, and EuroSys used PageRank as a benchmark.

Similarly, there is also a preference exhibited by the conference venue towards
some datasets as can be seen in \autoref{fig:dist:dataset_conf}. Papers
accepted at KDD tend to overuse the Twitter2010~\cite{twitter2010} and
yahoo-webscope datasets. 55\% of the papers at OSDI use the
uk-2007~\cite{uk200705} dataset, which is at least 40\% more than the usage of this
dataset in any other conference.

These results indicate the overuse of certain datasets and benchmarks and the
use of these together in preferred pairs.

\begin{table*}[ht]
\centering
\begin{tabular}{|l|l|r|r|r|r|}
\hline
Conference & Period  & \multicolumn{1}{l|}{Graph Papers} & \multicolumn{1}{l|}{\begin{tabular}[c]{@{}l@{}}Artifact\\ Available\end{tabular}} & \multicolumn{1}{l|}{\begin{tabular}[c]{@{}l@{}}Artifact \\ Evaluated\end{tabular}} & \multicolumn{1}{l|}{\begin{tabular}[c]{@{}l@{}}Result\\ Reproduced\end{tabular}} \\ \hline
SOSP       & 2019-21 & 3 (92)                            & 1 (59)                                                                            & 2 (46)                                                                             & 1 (39)                                                                           \\ \hline
OSDI       & 2020-23 & 1 (205)                           & 1 (135)                                                                           & 1 (131)                                                                            & 1 (113)                                                                          \\ \hline
EuroSys    & 2021-23 & 6 (137)                           & 2 (85)                                                                            & 2 (69)                                                                             & 0 (42)                                                                           \\ \hline
SIGMOD     & 2016-17 & 17 (341)                          & 0 (14)                                                                            & 0 (8)                                                                              & 0 (22)                                                                           \\ \hline
SIGMOD     & 2018-19 & 11 (385)                          & NA                                                                                & NA                                                                                 & 1 (17)                                                                           \\ \hline
SIGMOD     & 2020-22 & 22                                & 0 (54)                                                                            & 0 (55)                                                                             & 0 (54)                                                                           \\ \hline
VLDB       & 2018-19 & 16 (483)                          & NA                                                                                & NA                                                                                 & 0 (3)                                                                            \\ \hline
VLDB       & 2020-22 & 23 (676)                          & 8 (302)                                                                           & NA                                                                                 & 0 (6)                                                                            \\ \hline
\end{tabular}

\caption{Artifact evaluation badges awarded to graph processing systems at various conferences. NA indicates that the conference did not hand out that specific badge during the specified time period. Numbers in parentheses show the total numbers for the conference in the specified period. (Artifact Evaluated column indicates the Reusable badge for SIGMOD and the Functional badge for other conferences).}
    \label{tab:artifact}
\end{table*}

\subsubsection{Artifact evaluation of graph processing systems}

There has been a strong push for artifact availability and result reproduction in the database and systems community, and most conferences we surveyed have an artifact evaluation effort.
There are however shortcomings to this approach - not all artifacts can be made publicly available for legal reasons, and the evaluated artifacts and results are not a citable entity.
This problem is pervasive to all science, and the data science community has taken the lead in fixing the problem of dataset and code availability by creating repositories such as Zenodo~\cite{Zenodo} and Dataverse~\cite{Dataverse}.
The Data Management and Systems communities can follow their lead and recommend that published papers come with citable and accessible datasets and software, detailed usability instructions, and the environmental setup needed to reproduce results.

\autoref{tab:artifact} shows the artifact evaluation badges awarded to the papers in our corpus by various conferences. We consider only those time periods in which a conference ran an artifact evaluation procedure. Out of the 55 papers that could have obtained the ``Artifact Available'' badge, less than one-fourth (12) papers were awarded the badge. Shockingly, out of the 99 papers that could have obtained the ``Results Reproduced'' badge, only three papers were awarded the badge. Note that the badges are only awarded if the authors of a paper explicitly sign up for the artifact evaluation process. The absence of a badge does not necessarily mean that the paper's artifact might not be publicly available or that its results might not hold.
\section{Quantitative Study}
\label{sec:quant}

\begin{table}
    \centering
    \begin{tabular}{|l|l|}
        \hline
        Processing System & Version \\
        \hline
        \hline
        Galois~\cite{nguyen2013lightweight} & Commit 7f20a9131667 \\
        \hline
        GraphChi~\cite{kyrola2012graphchi} & Commit 6461c89f217f \\
        \hline
        Gemini~\cite{zhu2016gemini} & Commit 170e7d36794f\\
        \hline
        Ligra~\cite{shun2013ligra} & Commit 64d7ef450a22\\
        \hline
    \end{tabular}
    \caption{Versions of Graph Processing Systems used in experiments}
    \label{tab:setup}
\end{table}

The previous section described many of the problems we encountered in the
empirical work on graph processing systems. This section provides a quantitative
demonstration of the severity of these problems. Specifically, we focus on
performance discrepancies caused due to vertex isomorphisms (i.e., different
vertex ID assignment) and the presence of zero degree vertices.

For our quantitative study, we chose four popular graph processing systems -
GraphChi~\cite{kyrola2012graphchi}, Ligra~\cite{shun2013ligra},
Galois~\cite{nguyen2013lightweight}, and Gemini~\cite{zhu2016gemini}. We chose
these systems as they are popularly used and cited and have well-documented
open-source implementations available for a wide variety of benchmarks.
\autoref{tab:setup} shows
the versions of the graph processing systems used in our experiments.

All of the results reported for GraphChi, Galois, and Gemini were collected on
an Intel i7-core 3.1GHz processor machine with 32GB of RAM. All of the results
reported for Ligra were collected on an Intel(R) Xeon(R) 8-core CPU E5-2407
v2@2.4 GHz processor with 98GB of RAM, due to Ligra's higher memory requirements.
\subsection{Effect of Vertex Orderings}
\label{sec:isomorphism}

Many graph processing systems, quite reasonably, process vertices in order of
their unique IDs. However, for any graph, there exist many
isomorphic graphs that can be obtained simply by renumbering the vertices. We
define a \textbf{Vertex Ordering} of a graph dataset as one particular
assignment of vertex IDs to the vertices in the graph.  We demonstrate that this is primarily due to the fact that different
Vertex Orderings produce different cache behavior, which in turn, leads to
sometimes significant performance differences.

Reordering graphs is not limited to renumbering vertices as different orderings of edges can also impact performance. In prior work, McSherry
et al~\cite{mcsherry2015scalability} show that ``Hilbert Ordering'' of the edges produces better performance results than the default order of most
datasets.
The effects of reordering graphs are so profound and surprising that there is active research into developing lightweight and efficient re-ordering schemes to speed-up graph processing~\cite{8573478, 10.1145/2882903.2915220, 7515998, zhang2016optimizing,hotz2019experiences}.

We demonstrate the performance impact of vertex orderings only; McSherry et al~\cite{mcsherry2015scalability} already demonstrated the impact of edge orderings.
We describe a small subset of the various Vertex Orderings that can be used in
benchmarking graph processing systems. We choose these Vertex Orderings as they are
either popular or have well-defined semantics that provide some robustness
guarantees. The Vertex Orderings are as follows:

\begin{itemize}
    \item \textbf{Default}: Ordering that the dataset ``ships'' with. We find
	that this \emph{default} ordering tends to be either overly optimistic or
	overly pessimistic in terms of cache behavior.
    \item \textbf{Degree}: Ordering where the nodes are assigned IDs in
    ascending order of their degree.
    \item \textbf{Revdegree}: Ordering where the nodes are assigned IDs in
    descending order of their degree.

\end{itemize}

\subsubsection{Effect on Benchmark Performance}

\begin{figure*}
    \begin{subfigure}{0.49\textwidth}
        \centering
        \includegraphics[width=\textwidth]{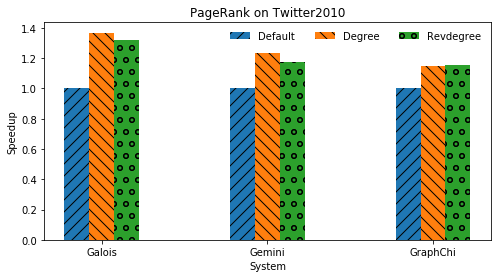}
        \caption{}
    \end{subfigure}
    \begin{subfigure}{0.49\textwidth}
        \centering
        \includegraphics[width=\textwidth]{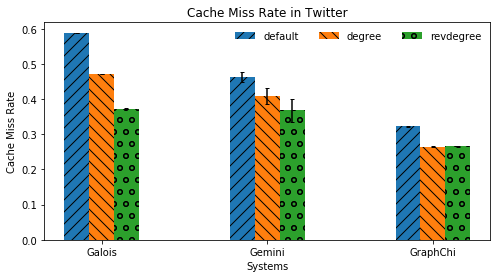}
        \caption{}
    \end{subfigure}
    \begin{subfigure}{0.49\textwidth}
        \centering
        \includegraphics[width=\textwidth]{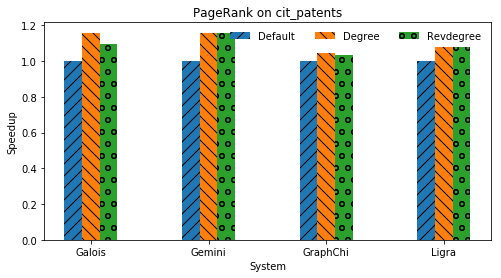}
        \caption{}
    \end{subfigure}
    \begin{subfigure}{0.49\textwidth}
        \centering
        \includegraphics[width=\textwidth]{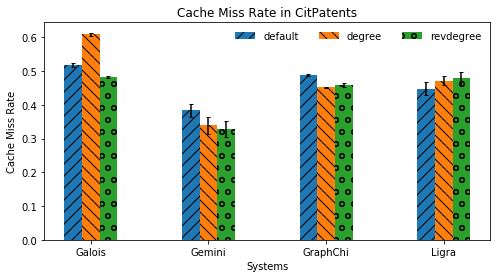}
        \caption{}
    \end{subfigure}
    \begin{subfigure}{0.49\textwidth}
        \centering
        \includegraphics[width=\textwidth]{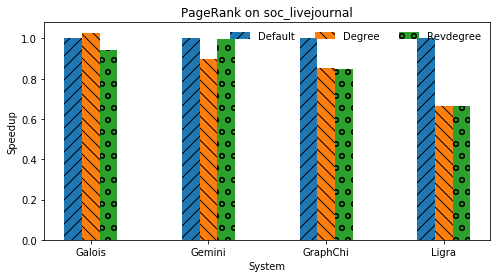}
        \caption{}
    \end{subfigure}
    \begin{subfigure}{0.49\textwidth}
        \centering
        \includegraphics[width=\textwidth]{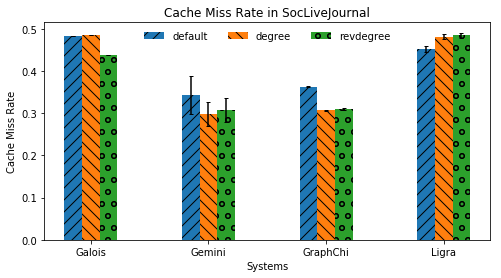}
        \caption{}
    \end{subfigure}
    \caption{Speedup attained on PageRank on Twitter2010, citPatents, and
    soc-LiveJournal with Galois, Gemini, GraphChi, and Ligra for Degree and Revdegree orderings as compared to the Default Ordering. We do not present
    results for running Ligra's PageRank on Twitter2010 as it did not finish
    within a 7-day time period.}
    \label{fig:pr_cache}
\end{figure*}

\begin{figure*}[h]
    \begin{subfigure}{0.49\textwidth}
        \centering
        \includegraphics[width=\textwidth]{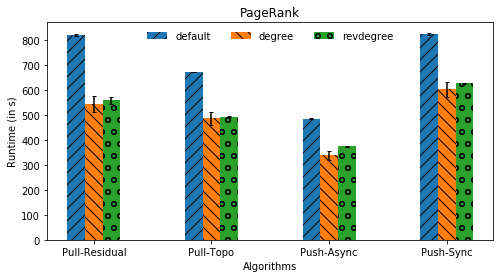}
        \caption{}
    \end{subfigure}
    \begin{subfigure}{0.49\textwidth}
        \centering
        \includegraphics[width=\textwidth]{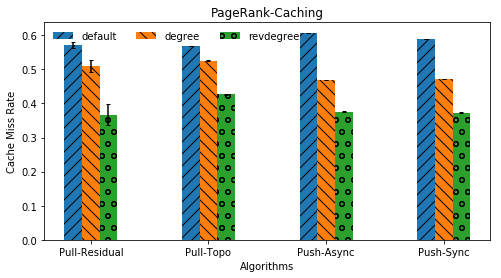}
        \caption{}
    \end{subfigure}
    \begin{subfigure}{0.49\textwidth}
        \centering
        \includegraphics[width=\textwidth]{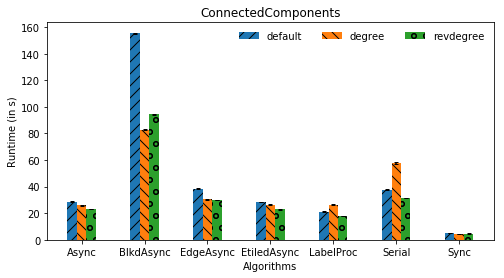}
        \caption{}
    \end{subfigure}
    \begin{subfigure}{0.49\textwidth}
        \centering
        \includegraphics[width=\textwidth]{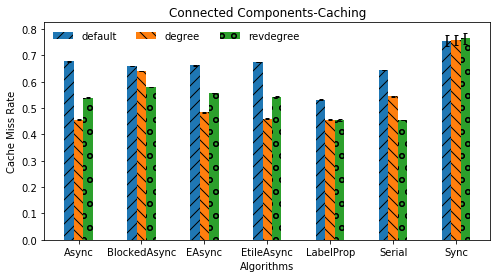}
        \caption{}
    \end{subfigure}
    \caption{Effect of Vertex Orderings on performance of various algorithms for
     running PageRank and ConnectedComponents with Galois on Twitter2s010 dataset.}
    \label{fig:alg-iso}
\end{figure*}

We demonstrate the effect of vertex order on runtime and cache behavior using
the most widely used benchmarks (Pagerank until convergence) on the most widely
used dataset (Twitter2010~\cite{twitter2010}), using three different Vertex Orderings (default,
degree, and revdegree) across different graph processing systems.
We do not
present any results for Ligra on Twitter2010~\cite{twitter2010},
as none of the runs completed after one full week.

To ensure a fair comparison between single-threaded and multi-threaded
graph-processing frameworks, we report runtime as the total ``user'' time across
all threads. We run our experiments on a hot cache, to produce accurate cache
measurements. We warm the cache by completing one run of the benchmark on the
dataset. We eliminate the effect of random noise by running each benchmark on
each isomorphism 25 times and report the mean and standard deviation for the
runtime and the cache miss rate.

\autoref{fig:pr_cache} shows the results of running different implementations
of PageRank algorithm with Galois, Gemini, and GraphChi on the Twitter2010~\cite{twitter2010},
citPatents~\cite{citPatents}, and soc-LiveJournal~\cite{socLiveJournal} datasets. 
We found that Degree Ordering seems to be
the best ordering across the graph processing systems for the Twitter2010~\cite{twitter2010}
dataset. However, it seems that the best ordering for other
datasets varies across systems. These results also illustrate that benchmark
performance is correlated with the cache behavior of the vertex ordering. In
our experiment, Degree Ordering for Twitter2010~\cite{twitter2010} yields a speedup of nearly
40\% compared to Default Ordering. This suggests that researchers must be
careful when comparing results across systems for any dataset, as the choice of
different orderings for the same dataset can produce significant, yet
misleading, results.

The difference in performance due to Vertex Orderings is not just present across
systems. Even in the same system, various implementations of the same algorithm
can produce drastically different results on the same dataset. This issue is
even more complicated, because different orderings produce best results for
different algorithms. \autoref{fig:alg-iso} shows the runtime for different
vertex orderings for all the different implementations of PageRank and
ConnectedComponents in Galois. We show the results for Galois as it has the greatest
number of different implementations for each benchmark. In a Bulk Synchronous
Parallel (BSP) system, a thread computing RageRank, for example, can "Pull"
updates from it's neighbors, or "Push" updates to it's neighbors
~\cite{whang2015scalable}. According to the Galois manual, ``the pull variants
perform better than the push. The residual version performs and scales the best
since it is optimized for improved locality and use of memory bandwidth". The
official Galois documentation and papers \cite{galois:tutorial} provide more
information about these algorithms and scheduling strategies.

Vertex Orderings can induce significant runtime differences.
In particular, default orderings are frequently overly optimistic or
overly pessimistic, so we should understand the default ordering as well as the impact that it can have on the system. 
\emph{In particular, it is essential that researchers avoid
inadvertently using different orderings on different systems.}

\subsubsection{Effect on Benchmark Correctness}
\label{benchmark_correctness}
\begin{table*}
    \centering
    \begin{tabular}{|l|r|r|r|}
        \hline
        Source & citPatents~\cite{citPatents} & soc-LiveJournal~\cite{socLiveJournal} & Twitter2010~\cite{twitter2010}\\
        \hline
        \hline
        GT: Cycle Triangles & 0 & 234455819 & 44738118599\\
        \hline
        GT: Trust Triangles & 7515027 & 946400853 & 143011093363\\
        \hline
        GT: Cycle + Trust Triangles & 7515027 & 1180856672 & 187749211962\\
        \hline
        Galois (NI) & 264897 & 132859748 & 22849431223\\
        \hline
        Galois (EI) & 188649 & 122558230 & 23027408359\\
        \hline
        Ligra & 7514962 & 187551052 & DNF\\
        \hline
        GraphChi & 7515023 & 285730264 & 34824916864\\
        \hline
        Snap Website & 7515023 & 285730264 & NA\\
        \hline
    \end{tabular}
    \caption{Number of Triangles reported by various graph systems for popular
    datasets. Rows marked with GT indicate the ground truth we calculated using the queries defined in \autoref{fig:triangle_alg}. DNF indicates that the benchmark did not finish in time. NA indicates that
    number of triangles was not available for a particular dataset.}
    \label{tab:triangle}
\end{table*}

Theoretically, the various vertex orderings should have no impact
on the correctness of the algorithm. However, we found that
for certain systems, the various Vertex Orderings can produce different
\emph{incorrect results} for triangle counting. Triangle Counting on
Ligra produces different numbers of triangles for the same
directed graph with different Vertex Orderings. On further inspection,
we found that this is a symptom of a larger problem:
there is no standard approach to
triangle counting in directed graphs.

\begin{figure*}[t]
\centering
\begin{subfigure}[b]{0.40\linewidth}
  \centering
  \begin{lstlisting}[
        language=SQL,
        showspaces=false,
        basicstyle=\tiny\ttfamily,
        numberstyle=\footnotesize,
        commentstyle=\color{gray}
     ]
    --Trust Triangle Count
    SELECT COUNT(*)
    FROM edges e1
    JOIN edges e2 ON e1.dst = e2.src 
    JOIN edges e3 ON e2.dst = e3.dst 
                 AND e3.src = e1.src 

                 
    \end{lstlisting}
\end{subfigure}
\begin{subfigure}[b]{0.40\linewidth}
  \centering
  \begin{lstlisting}[
        language=SQL,
        showspaces=false,
        basicstyle=\tiny\ttfamily,
        numberstyle=\footnotesize,
        commentstyle=\color{gray}
     ]
    --Cycle Triangle Count
    SELECT COUNT(*)
    FROM edges e1
    JOIN edges e2 ON e1.dst = e2.src 
                 AND e1.src < e2.src
    JOIN edges e3 ON e2.dst = e3.src
                 AND e3.dst = e1.src
                 AND e2.src < e3.src;
    \end{lstlisting}
\end{subfigure}
\caption{SQL queries to compute two different types of triangles for directed graphs}
    \label{fig:triangle_alg}
\end{figure*}

We found at least two definitions of triangles on directed graphs: cycle
triangles and trust triangles (when two connected vertices both have directed
edges to a third common vertex)~\cite{santoso2018triangle}.
\autoref{fig:triangle_alg} shows two SQL queries that compute the number of trust and cycle
triangles on a graph represented by an edge table, where each tuple in the
edge table represents a directed edge.
We compute the ``ground truth'' number of triangles
in a given dataset using these queries and compare the results to those
obtained from a brute
force implementation that runs over a text-based, edge-list representation,
common to many published graph datasets.
The bottom two rows of \autoref{tab:triangle} show
the number of cycle and trust triangles obtained from running these queries
on citPatents~\cite{citPatents},
soc-LiveJournal~\cite{socLiveJournal}, and Twitter2010~\cite{twitter2010}. 

However, we found that popular graph processing systems do not count cyclical or
trust triangles correctly on directed graphs. We found only one
paper~\cite{hong2013early} that specifies counting trust triangles for their
triangle counting benchmark on directed graphs. Disturbingly, triangle counting
implementations of these systems differ sufficiently that running them on
identical datasets produces different numbers of triangles, as shown in
\autoref{tab:triangle}. Galois's triangle counting benchmark assumes that the
graph is undirected, and when run on directed graphs, the benchmark's behavior
is essentially undefined. This undefined behavior is elegantly depicted by the
fact that Galois' two implementations of triangle counting, Node Iterator and
Edge Iterator, report different numbers of triangles. Ligra enforces no such
restriction and measures trust triangles. Unfortunately, their
implementation depends on vertex ID order. (A trust triangle is counted
only if the directed edges go from a higher numbered vertex ID to a lower
numbered vertex ID.) Thus, two graphs that are identical, except for vertex
relabeling (that is, two isomorphic graphs), will yield different triangle
counts. Note that for undirected graphs, the different systems report the same
number of triangles.

Additionally, one paper in the corpus uses an approximate triangle counting algorithm
based on top-n eigenvalues called EigenTriangle~\cite{tsourakakis2008fast}. As this is an approximate
algorithm, it will not yield 100\% accurate results all the time. We have no insight as to how many other papers
in our corpus might have used this or another approximate algorithm as the use of approximate algorithms is not documented in these papers.
\subsection{Effect of Zero-Degree Vertices}
\label{sec:zero}

\begin{table}[h]
\centering
\resizebox{\columnwidth}{!}{%
\begin{tabular}{|l|r|r|r|r|}
\hline
\textbf{Dataset} &
  \multicolumn{1}{l|}{\textbf{\begin{tabular}[c]{@{}l@{}}Total Vertices\end{tabular}}} &
  \multicolumn{1}{l|}{\textbf{\begin{tabular}[c]{@{}l@{}}Zero-Degree Vertices\end{tabular}}} &
  \multicolumn{1}{l|}{\textbf{\begin{tabular}[c]{@{}l@{}}Non Zero-Degree Vertices\end{tabular}}} &
  \multicolumn{1}{l|}{\textbf{\begin{tabular}[c]{@{}l@{}}\% of Zero-Degree Vertices\end{tabular}}} \\ \hline
Twitter2010~\cite{twitter2010} & 61578414 & 19926184 & 41652230 & 32 \\ \hline
citPatents~\cite{citPatents} & 6009554 & 2234786 & 3774768 & 37 \\ \hline
soc-LiveJournal~\cite{socLiveJournal} & 4847570 & 0 &  4847570 & 0 \\ \hline
friendster~\cite{friendster} & 124836179 & 59227813 & 65608366 & 47 \\ \hline
uk-2007-05~\cite{uk200705} & 105896434 & 677865 & 105218569 & 6 \\ \hline
\end{tabular}%
}
\caption{Total number of vertices and zero-degree vertices in popular datasets}
\label{tab:zerodeg}
\end{table}

Zero-degree vertices, or isolated vertices, have neither incoming nor outgoing edges.
In graph datasets, these vertices do not appear in edge lists, but their IDs lie within the range of the minimum and maximum vertex IDs.
\autoref{tab:zerodeg} shows the total number of vertices and the number of zero-degree vertices of popular datasets.
The large number of zero-degree vertices in the Friendster dataset~\cite{friendster} explains why some describe the dataset as a 124 million node graph~\cite{lee2017extrav}, while others describe it as a 65 million node graph~\cite{roy2013x}.
Zero-degree vertices affect benchmark results in myriad ways.

\subsubsection{BFS and Zero-Degree Vertices}

\begin{figure}[h!]
    \centering
    \includegraphics[width=0.48\textwidth]{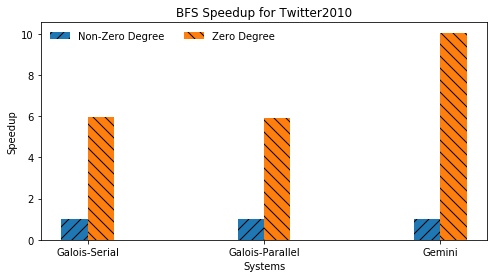}
    \caption{BFS performance comparison between starting with a zero-degree vertex and a non-zero-degree vertex.}
    \label{fig:zero_bfs}
\end{figure}

As mentioned in \autoref{sec:isomorphism}, the BFS ordering is ill-defined,
because many real-world graphs do not have a natural root node. In the absence
of any natural root, many graph processing
systems~\cite{shun2013ligra,zhu2016gemini,nguyen2013lightweight} use  the vertex
with ID 0 as the default root and vertex with ID 1 as the target node. However,
chaos ensues in some benchmarks if that first vertex happens to be zero-degree
 node, as is the case in three of the five 
datasets in \autoref{tab:zerodeg}.

\autoref{fig:zero_bfs} illustrates the effect of choosing
a zero-degree vertex versus a non-zero-degree vertex as the
starting node for the BFS benchmark on Galois and Gemini.
We randomly select 20 different non-zero-degree vertices from the
Twitter2010~\cite{twitter2010} dataset and measure BFS runtime starting from that node
on Galois's serial-mode and parallel-mode BFS and on Gemini.
For Gemini, we use the authors' built-in measurement mechanism, which
measures the end-to-end runtime and not the time spent by all threads.
As \autoref{fig:zero_bfs} shows, choosing a zero-degree vertex on
Twitter2010~\cite{twitter2010} produces a 6x-7x speedup
on Galois and nearly a 10x speedup on Gemini.

As most reported results in the literature fail to report either their
Vertex Ordering or their BFS start node, it is virtually impossible
to extract meaning from such results.
\ms{It would be super great if we could claim the following: }
Based on our experiments, we hypothesize that the BFS results shown in the
comparison of Galois and Ligra in the original Galois
paper~\cite{nguyen2013lightweight} are using zero-degree BFS time as both Ligra
and Galois have the same runtime on Twitter2010~\cite{twitter2010} which, in our experience,
happens only if they are using the vertex with ID 0 as the root node, which the
implementations of both these systems indeed do by default.

\subsubsection{Effect on Triangle Counting}

\begin{figure}[h!]
    \centering
    \includegraphics[width=0.48\textwidth]{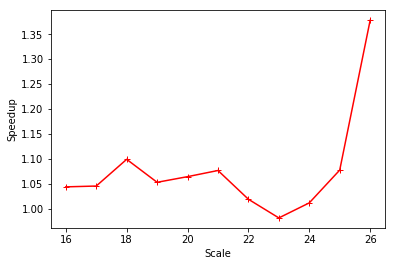}
    \caption{Speedup obtained on Triangle Counting with GraphChi after removing
    zero-degree vertices from RMAT generated graphs with scale ranging from 16
    to 26.}
    \label{lab:zero_rmat}
\end{figure}

While zero-degree vertices are most problematic for benchmarks that use a start
node, they also have an effect on other benchmarks. To better understand the
effect of removing zero-degree vertices, we generated 11 RMAT graphs with
``scale'' ranging from 16 to 26 with an ``edgefactor'' of 26. In an RMAT graph,
$|V| = 2^{scale}$ and $|E| = |V| \times \texttt{edgefactor}$. However, these
vertices also include zero-degree vertices, and increasing scale increases both the absolute number and proportion of zero-degree vertices.
\autoref{lab:zero_rmat} shows the effect of removing
zero-degree vertices when running Triangle Counting on GraphChi.
Increasing the scale factor (X-axis) increases both the ``size of the graph''
and the fraction of zero-degree vertices.
\autoref{lab:zero_rmat} shows that the speedup obtained from removing these
zero-degree vertices seems to grow exponentially once the graph becomes
sufficiently large (i.e., at scale=23).
This is a concern, considering that in the absence of real data for large graphs,
researchers increasingly rely on synthetic generators. If a
particular system were to remove the zero-degree vertices
during preprocessing, the system could potentially produce speedups
of more than 30\%.

\subsubsection{Effect on PageRank and ConnectedComponents}

\begin{figure*}[hbt]
    \begin{subfigure}{0.48\textwidth}
        \centering
        \includegraphics[width=\textwidth]{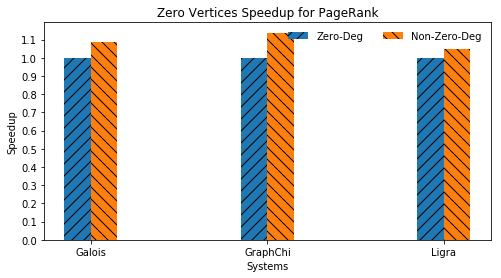}
        \caption{}
    \end{subfigure}
    \begin{subfigure}{0.48\textwidth}
        \centering
        \includegraphics[width=\textwidth]{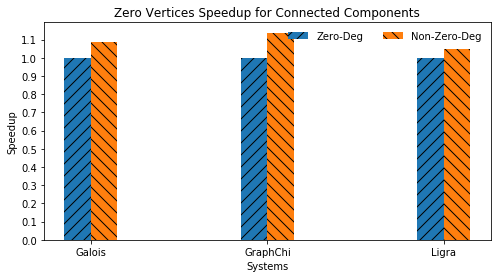}
        \caption{}
    \end{subfigure}
    \caption{Effect of removing zero-degree Vertices from citPatents Dataset; 
    for running PageRank benchmark using Galois, GraphChi, and Ligra}
    \label{lab:zero_pr}
\end{figure*}

Removal of zero-degree vertices can also produce speed-up for PageRank and Connected Components. \autoref{lab:zero_pr} shows the speedup gained on PageRank
and Connected Components by removing
zero-degree vertices from citPatents~\cite{citPatents} in Galois, GraphChi, and Gemini. We use the Push-Sync
algorithm for PageRank, and we measure the time until convergence. Notice that
for both benchmarks, removing zero-degree vertices can produce a speedup of up
to 13\%. Thus, we would advise that researchers be wary of the impact of
zero-degree vertices when evaluating systems on different benchmarks. 
\section{Best Practices}
\label{sec:best}

Based on our findings from the literature study in \autoref{sec:metastudy}
and impact of dataset properties on benchmark performance in \autoref{sec:quant},
we propose a set of best practices for
benchmarking graph processing systems.
\\\\
\textbf{Standardization:} To overcome the lack of diversity in the datasets and benchmarks used, the community should develop a rigorously specified set of standardized benchmarks and should publish more well-described, large datasets. A rigorous benchmarking approach would be to use a mix of real-world and (well-constructed) synthetic datasets of varying topologies. As identified by Besta et. al.\cite{10.14778/3476249.3476252}, it is critical to use datasets of different origins (e.g., road networks, social networks, biological networks, etc.) to evaluate graph processing systems. The current practice of poorly matched dataset-benchmark combinations will undoubtedly propagate to future research through citation networks, unless we promote constructive diversity in benchmarking practices.
\\\\
\textbf{Synthetic Graph Generators:} For generating large graphs for evaluating the scalability of graph processing systems, researchers should use the Smooth Kronecker graph generator~\cite{anand2020smooth} instead of other RMAT or Kronecker graph generators.
\\\\
\textbf{Dataset Hashes:} Dataset providers should publish hashes of their datasets, and authors should report them. In addition to publishing and citing hashes, the community needs to use standard names and DOIs.
\\\\
\textbf{Detailed Metrics:} As recommended in GAIL~\cite{10.1145/2833179.2833187}, benchmarking efforts and reporting should be expanded to include more detailed metrics about kernel execution time, the amount of algorithmic work (number of edges traversed per kernel), cache effectiveness (the number of memory requests made per edge traversed), and the memory bandwidth utilization (time taken per memory request).
\\\\
\textbf{Preprocessing:} Preprocessing the datasets constitutes a significant part of the overall runtime of graph processing systems. The preprocessing time can often be amortized if the same data format is used for several tasks and if the graph doesn't evolve too rapidly. However, it is critical to measure and report the preprocessing overheads so that users can make an informed choice about how well-suited the system is to their workload and dataset.\\
Additionally, it is vital to report what preprocessing steps a system undertakes (removing zero-degree vertices, for instance). Reporting the preprocessing steps carried out of a dataset and the time these steps take must be incorporated into the benchmarking suites and general best practices.
\\\\
\textbf{PageRank Usage:} When using the PageRank benchmark, there are two ways to report the system's performance: the time taken for the PageRank values to converge to a tolerance value or the time taken to complete a fixed number of iterations. We recommend using the time to converge to a tolerance as the reported PageRank performance.
Time to convergence is a more robust metric because it is meaningful for both synchronous and asynchronous modes of computation~\cite{sync_or_async}
It is also more widely used with newer algorithms that track active vertices in each iteration.
\\
However, these metrics alone cannot explain which choices (the system itself, the PR algorithm being used, or the hardware) most impact performance.
If an optimized version of PR is used, for example, one that maintains active vertices for each iteration, the runtime indicates the choice of a superior implementation as much as the system's ability to traverse the graph efficiently. 
Reporting the convergence time alone conceals that the algorithm traverses fewer edges with each iteration.
The performance measurement would be more robust when coupled with information about the total number of edges the algorithm traverses.
Traversed Edges Per Second (TEPS) alone is not sufficient and can be misleading, so ideally a deeper measurement of the kernel performance should be performed, as detailed in section ~\autoref{subsec:TEPS}
\\\\
\textbf{Benchmark \& Dataset Selection:} Benchmark selection and Dataset selection are not separable; researchers should use datasets appropriate for the benchmark or the system property (scalability, for instance) that needs to be elided.
While there are no strict rules defining what datasets are "appropriate" for a specific benchmark, researchers should aim to incorporate a variety of datasets that challenge their system and reveal potential bottlenecks.
While large datasets are frequently selected to showcase scalability, merely assessing graph size is inadequate for predicting its effectiveness in stressing the system.
Other factors like degree skewness, sparsity, diameters, and locality (defined as preferred) are valuable in assessing system performance, and therefore, the chosen datasets should encompass a broad spectrum of these characteristics.
\\\\
\textbf{Triangle Counting:} Researchers who use triangle counting as a benchmark must specify A) whether the graph is directed or undirected, B) if directed, whether they are counting cyclical triangles, trust triangles, or both, and C) the algorithm being used and whether it computes exact or approximate triangle counts.
An exact triangle counting implementation should never return different counts depending on the IDs assigned to vertices. We encourage the community to adopt trust triangles as the official definition of a triangle in directed graphs, as the definition of a trust triangle is entirely independent of the dataset's vertex ordering.
\\\\
\textbf{Vertex Orderings:} To nullify the impact of different vertex orderings, researchers must report the ordering used.
We recommend that researchers show results for at least
three different orderings: default (i.e. the same ordering as on the data source), Degree, and RevDegree. 
Under no circumstances should researchers use different orderings of the same dataset for different systems. 
\\\\
\textbf{Correct Start Nodes:} Any benchmark whose results depend on a start node should ensure that their start nodes are not zero-degree and report the start node(s) used unambiguously.
\\\\
\textbf{Handling Zero-Degree Vertices:} To deal with the hidden effects of zero-degree vertices, researchers should use packed representations that remove zero-degree vertices from datasets. Additionally, researchers should explicitly state the total number of vertices and the number of zero-degree vertices in their datasets.
\\\\
\textbf{Artifact Availability:} Authors of graph processing papers should, at the very least, make their artifacts publicly available via the Artifact Evaluation procedure. As part of the artifact, they must include their system, all preprocessing steps, and DOIs to datasets they used as part of their evaluation.
\section{Related Work}\label{related_work_section}
One of the earliest warnings about improper use of datasets and benchmarks comes from  Sebastiano Vigna~\cite{vigna2007stanford} when he discussed how using PageRank as a benchmark is ``interesting'' only on \emph{``large''} datasets. 
Since then, it has become common knowledge that benchmarking large graph processing systems is difficult.
This is best captured in the LDBC vision paper \cite{guo2014benchmarking}: 
The authors describe the challenges that arise from the diversity in datasets, algorithms, and platforms used in graph benchmarking.
Owing to this difficulty in gaining clear insights into the system and its performance, several attempts have been made to compare different graph processing frameworks in a more standardized setup. 
Experimental studies, such as the one by Nadathur et al. \cite{satish2014navigating}, compare graph processing systems precisely and shed much-needed light on the performance characteristics of these systems.
The authors first establish benchmark baselines using hand-optimized code and then compare the performance of graph frameworks on these benchmarks.
They use it to ``delineate bottlenecks arising from algorithms themselves vs. programming model abstractions vs. the framework implementations.''

The remainder of this section examines important prior work that called for better benchmarking practices, often suggesting new metrics and frameworks to do so and why academia remains stubbornly opposed to using them.

\subsection{COST: Overreliance on Scalability}
\label{subsec:COST}
The strong scalability~\cite{10.5555/2385466} of a system is defined as the ratio $T_1/T_N$, with $T_1$ being the time taken to complete a task with one thread and $T_N$ being the time taken with N threads.
Scalability has been an important metric to measure performance improvement as more cores become available, and consequently, is widely used to measure distributed graph processing systems.
However, only observing the scalability of a system against that of other graph processing systems fails to identify the unclaimed improvements that could be realized with better optimizations.

This is the premise of the COST metric \cite{mcsherry2015scalability}, which can be used to identify how close a graph processing system performs to a reference implementation that is hand-optimized and single-threaded.
COST is the number of threads a system must use to perform better than this reference implementation; according to the authors, several distributed systems had an unbounded COST (never outperforming the reference implementation).
While COST does not represent the lowest possible time needed for a task (it could be reduced further with the use of more modern algorithms, for instance), it does provide an important yardstick to assess the ``optimization level'' of a system -- the closer to 1, the closer the system is to a performant baseline.

Despite being an easy, reliable, and revealing metric, only a handful of research papers have followed the advice of the COST paper and demonstrated scaling results against a single-thread configuration.
Gemini~\cite{zhu2016gemini}, Automine~\cite{automine_sosp19}, GRAM~\cite{Gram_Socc15}, G-POP~\cite{gpop_2020} are few notable exceptions. 

The COST paper highlights the importance of better baselines, but more importantly, also highlights how hard it is to compare systems and the need to ``always measure one level deeper''~\cite{10.1145/3213770}.
Our paper continues the COST work by identifying and rectifying the most pervasive forms of incorrect benchmarking practices prevalent in the community. 

\subsection{Traversed Edges per Second can be Misleading}
\label{subsec:TEPS}
The performance of a system is dependent on several factors, and the measurement of execution time is not sufficient to glean any insights into \textit{why} the system performs the way it does.
Another metric commonly used to measure the performance of graph processing systems is TEPS: the number of edges the system can scan per second.
This is a better metric and provides some insight into the system that is independent of the dataset used for the benchmark.
Beamer et al.~\cite{10.1145/2833179.2833187} note that TEPS can be misleading if the counting of an edge is not fixed (undirected edges are often represented as two directed edges -- should we count them once or twice?).
More importantly, however, a better algorithm can reduce the number of edges that need to be scanned, leading to a lower TEPS score but faster execution.
Instead, they propose the Graph Iron Law - GAIL, which allows performance engineers to "weigh the trade-offs between the three most important graph algorithm performance factors: algorithm, implementation, and hardware platform."
GAIL allows the decomposition of the execution time into a product of the number of edges an algorithm needs to traverse, the number of memory requests each traversed edge makes, and the time taken to service each such memory request:
\[
\small
    \dfrac{\text{time}}{\text{kernel}} = \text{ } \dfrac{\text{edges}}{\text{kernel}} \text{ x } \dfrac{\text{memory requests}}{\text{edge}}  \text{ x }  \dfrac{\text{seconds}}{\text{memory requests}}
\]
While this metric has been designed for single-node shared memory systems, it can easily be extended for disk-based systems by measuring the number of read blocks and for distributed systems by measuring the number of packets exchanged.

GAIL remains underutilized in research, and we only managed to find four publications that directly reference it, yet several papers use TEPS as a primary metric to evaluate algorithmic throughput.

\subsection{Benchmark Suites: Standardization of Graph Benchmarking}

Other than the measurement problems mentioned in \autoref{subsec:COST} and \autoref{subsec:TEPS}, several other issues with common comparative benchmarking approaches need careful consideration.
Comparing one system to another is most meaningful only when done so that the difference between the systems is the novel contribution of the newer system. 
However, this is impossible and can rarely be achieved in practice.
More realistically, what can be controlled is the methodology of conducting the evaluation: collecting the same metrics on algorithms that do the same amounts of work when run on the same datasets.
Using a standard input dataset can also uncover spurious optimizations that hold only for some datasets but do not generalize to a diverse selection.
Most importantly, as Beamer et al. note, ``using a standard high-quality reference implementation could help discourage using low-performance baselines'' \cite{beamer2015gap}.

Thankfully, there has been a strong push towards standardization to address the broken benchmarking practices in the community.
Much research went into producing good benchmarks and identified datasets that can indicate the realistic performance of the systems being profiled.
Starting with Graph500 \cite{graph500}, there have been many efforts toward making the benchmarking process more streamlined. 
Graph500 defined benchmark kernels (BFS and SSSP), provided a synthetic graph generator, instructions on what metrics must be collected and reported, and a validation protocol for submitted results.
Graph500 provides only two kernels and their fixed-topology synthetic graph generator graphs, which exhibit unusual statistical characteristics, as mentioned in Section \ref{sec:zero}~\cite{anand2020smooth}.
While Graph500 is well-liked and useful for HPC and supercomputing benchmarking, these shortcomings make it less useful for benchmarking graph processing systems.

There have been several subsequent benchmarking suites; the survey by Bonifati et al. \cite{bonifati2018survey} describes early benchmark suites and their novel contributions, programming abstractions, datasets, and workloads.
The most recent and relevant benchmarking suites that we focus on in the remainder of this section are GAP-BS~\cite{beamer2015gap}, Graphalytics~\cite{ldbcsite}, and GraphMineSuite~\cite{10.14778/3476249.3476252}
Each of these graph benchmarking suites provides performant reference code, but Graphalytics, LDBC-SNB, and GraphMineSuite also provide a variety of datasets of different sizes and domains. We now look at them in more detail:

\textbf{GAPBS \cite{beamer2015gap}} attacks the spurious comparison problem by specifying the benchmarks that other system designers can use when implementing their own benchmarks to ensure that their system can be more widely compared to other specification-adherent systems.
They also provide a reference (shared-memory) implementation of six benchmark kernels: Pagerank (PR), Breadth-First Search (BFS), Single-Source Shortest Path (SSSP), Connected Components (CC), Betweenness Centrality (BC), and Triangle Counting (TC).
These can be used to establish a performant baseline against which to compare.
Each reference implementation also comes with a verifier function that can be used to verify the correctness of the solution by either testing for the properties of a correct solution (for PR, BFS, or CC) or running a dead-simple serial implementation of the benchmark kernel (for SSSP, BC, or TC).
The benchmark provides five input graphs: one uniform random  Erdős–Réyni synthetic graph, one Kronecker graph, and three real-world graphs (Twitter, Web sk-2005, and USA-Road).
GAP-BS is actively maintained and quite widely used.
The primary critique of GAPBS is that they do not provide more datasets, and the ones they do provide are rather small and not statistically diverse enough.

\textbf{LDBC Graphalytics~\cite{iosup2023ldbc}} is perhaps the most well-known graph processing benchmark suite, which was designed to bring order to the world of benchmarking graph analysis systems.
Graphalytics provides a specification and reference implementation of six algorithms: PR, BFS, Weakly Connected Components (WCC), Community Detection using Label Propagation (CDLP), SSSP, and Local Clustering Coefficient (LCC).
System developers can implement their algorithms as long as their correctness can be validated compared to a reference output.

Graphalytics has the same vision as GAPBS but differs from GAPBS in a few important ways. 
First, Graphalytics provides a more detailed breakdown of the end-to-end time the graph processing task takes. It provides the breakdown of the total runtime of the task by the time to load the graph, the time for setup and teardown operations, the time taken to preprocess and run Extract-Transform-Load operations, and the time to run the task itself.
Second, Graphalytics improves upon GAPBS by providing a richer set of metrics for expressing the processing throughput. Graphalytics reports the Traversed Edges per Second (TEPS) and reports Edges and Vertices per Second (EVPS) (the sum of the number of edges and the number of vertices processed by the system per unit time). They also provide two cost metrics that measure the Total Cost of Ownership (calculated based on the TPC pricing model) and Price Per Performance (defined as the ratio between TCO and EVPS).
Third, Graphalytics provides a larger set of real-world and synthetic datasets of various sizes and domains. They also provide several synthetic graphs generated by the Graph500 Kronecker generator and the LDBC Datagen tool. Graphalytics identifies the ground truth by generating a reference output using a ``specifically chosen reference platform, the implementation of which is cross-validated with at least two other platforms up to target scale L. The results are tested by cross-validating multiple platforms and implementations against each other''. Any output generated can be verified against this reference output by using either an exact match (for BFS and CDLP), equivalence matching (which applies to WCC), or an epsilon matching strategy (which applies to PR, LCC, and SSSP).

Graphalytics also provides a useful Java-based test harness and driver code that can be used to run benchmarks on a system in a manner consistent with their specification.
However, making use of this harness requires adapting their generic platform driver with the platform-specific implementation.
This is great for reliable and reproducible benchmarking practices; however, this is an almost insurmountable challenge to an academic researcher who is pressed for time and has little to no incentive to do this extra work.

\textbf{GraphMineSuite~\cite{10.14778/3476249.3476252}} is a recent benchmark suite that was designed to develop highly performant baselines for graph mining algorithms. 
GMS has been designed explicitly to handle graph problems that are \textit{not} traditionally researched in the parallel processing community (such as PR, BFS, etc.); instead, GMS chooses to focus on the equally important but harder problems of graph pattern matching (maximal cliques, k-clique, frequent subgraph mining, subgraph isomorphism, etc.), graph learning (vertex similarity, link prediction, clustering, and community detection), optimization problems (graph coloring, minimum spanning tree, minimum cuts), and problems of vertex reordering (degree reordering, triangle counting ranking, and degeneracy reordering).
In all, GMS provides some forty high-performant benchmark reference implementations and specifications.

GMS reports all the usual metrics that one expects from a benchmark suite: kernel running times and their breakdown by task, strong scaling analyses, memory consumption, machine efficiency (CPU core utilization), and memory bandwidth utilization (measured in terms of cache hits/misses, memory reads/writes, etc.).
GMS defines a new metric for measuring algorithmic throughput that ``measures the number of mined graph patterns per unit time''.
Depending on the algorithm domain, this metric measures different things: graph subgraphs found per second for graph mining problems, similar vertices derived per second for link prediction tasks, and the number of communities detected per second for clustering/community detection tasks.

GMS, unlike other benchmarking suites, defines guidelines to select datasets that would be computationally challenging for all specified algorithms in the suite.
While GAPBS chose input graphs of varying degrees of sparsities, diameters, degree distribution skew, and amount of locality,
GMS emphasizes including datasets from different origins, since they observed that two graphs of nearly identical statistical properties (size, densities, degree skew, and diameters) can behave differently for the same mining task, since the dataset's domain might make it likely to contain certain motifs. 
\\\\
These benchmark suites have helped advance and standardize benchmarking and bring some much-needed rigor to the area of research.
Our work shares the fundamental vision of better benchmarking practices proposed by all these benchmarking frameworks.
We focus on illustrating the significant consequences of subtle modifications to the dataset. 
Vertex ordering is an important factor in graph analysis and should be explicitly controlled for, as in the study by Abbas et al. \cite{abbas2018streaming}.
To our knowledge, vertex re-orderings are not explicitly controlled for in any comparative study we have encountered in our research corpus.
To the best of our understanding, we are the first to demonstrate the impact of isolated vertices in synthetic datasets on benchmark performance.

\subsection{Lack of Adoption in Academia}
Benchmarking suites are great, yet they remain pitifully underutilized in academia. Why?
Consider the composition of the LDBC organization~\cite{tpctc_ldbc_organization}: they have 24 member organizations out of which only 3 are non-commercial.
LDBC also runs a tournament-style competition of graph processing systems~\cite{musaafir2018specification}.
Of the recent submissions to the Graphalytics competitions - nearly all are from the industry, which is understandable considering that an audited performance profile is a great marketing tool. 
Academics usually do not have any similar incentive.
Researchers need to benchmark their system against any system they identify as a close competitor, and they likely need to do so quickly before an impending paper submission deadline.
Using graphalytics would be a herculean effort as we explain below:

The Graphalytics test driver is written in Java and one needs to implement the relevant classes to make use of it. 
This effort must be repeated for each system under test. 
This is not likely to happen in academia, where we are only now catching up to the practice of artifact evaluation, sharing citeable datasets and software versions, and automatically reproducible results.
Graphalytics presents an all-or-nothing proposition for academic research: either everybody uses it and benchmarking is utopian, or nobody does, and we remain in the benchmarking rut we currently find ourselves in.
Ideally, academics should use Graphalytics for benchmarking (and the artifact evaluation committees ought to enforce that), but if they choose not to, they should strive to be more proactive in reporting additional metrics and implementation details.

A more attainable goal would be to have graph processing systems explicitly state which graph algorithm they use for their benchmark kernel implementation and whether their implementation conforms to the specification laid out by a benchmarking suite such as GAPBS, GBBS, or GMS.
Artifact evaluation committees should mandate this in all major data management conferences.
Similarly, our benchmarking best practices are easy to enforce and do not require much effort from academics. 
\section{Conclusion}

Graph-structured data is increasingly important, and we see no indication that
further work in producing efficient and scalable graph processing systems will
cease. However, ensuring that next-generation graph processing systems are truly
advancing the state of the art requires that we use sound methodology in
assessing performance. We have shown that current practice falls far short of
this ideal. It might be tempting to interpret the lack of diversity in graph data
sets and benchmarks as a kind of standardization. However, the majority of these
data sets are too small to demand the scale of modern processing systems; they
fit in the main memory of most laptop and desktop machines.

State-of-the-art evaluations suffer from poor evaluation practices, such as imprecisely describing datasets, failing to both describe and measure preprocessing steps, omitting crucial experimental setup parameters, and ignoring zero-degree vertices and vertex orderings.
This last omission is particularly troubling. 
Different vertex orderings lead to drastically different results for the same
graph on the same graph processing system, yet most papers are silent on such
orderings. Zero-degree vertices can render BFS benchmark results meaningless.
Every graph processing system seems to pick a unique definition of a
``triangle'' when it comes to the Triangle Counting benchmark on directed graphs. Good science requires that we do better.

In this SoK, we present a review of the benchmarking practices used in academic research on graph processing systems and highlight the pressing issues that need to be addressed for codification and standardization.
We demonstrate the impact of underspecifying the datasets, their properties, and the preprocessing steps performed during the evaluation and provide recommendations for reporting and addressing these issues.
The SoK details previous attempts at benchmarking suites that partly address these problems and provide actionable insights for researchers to incorporate into their benchmarking practices.
We believe this systematization and our recommendations will guide future research efforts and ultimately improve the reproducibility and robustness of published results.

\balance


\bibliographystyle{unsrt}
\bibliography{paper}


\end{document}